\documentclass[aps,prl,twocolumn,nofootinbib,tightenlines,showpacs,%
showkeys,floatfix,tightenlines,preprintnumbers]{revtex4}

\usepackage{epsfig,bm}

\begin{document}

%\preprint{hep-ph/0402135}

%%%%%%%%%%%%%%%%%%%%% Title %%%%%%%%%%%%%%%%%%%%%%

\title{Anti-charmed pentaquark from $B$ decays}

%%%%%%%%%%%%%%%%%%%% Authors %%%%%%%%%%%%%%%%%%%%%
\author{Su Houng Lee}%
\email{suhoung@phya.yonsei.ac.kr}

\author{Youngshin Kwon}
\email{kyshin@yonsei.ac.kr}

\author{Youngjoon Kwon}%
\email{kwon@phya.yonsei.ac.kr}

\affiliation{Institute of Physics and Applied Physics,
Yonsei University, Seoul 120-749, Korea}

%%%%%%%%%%%%%%%%%%%% Abstract %%%%%%%%%%%%%%%%%%%%%

\begin{abstract}
  We explore the possibility of observing the anti-charmed pentaquark
  state from the $\Theta_c \bar{n} \pi^+$ decay of $B$ meson produced
  at $B$-factory experiments.  We first show that the observed
  branching ratio of the $B^+$ to $ \Lambda^-_c p \pi^+$, as well as its open
  histograms,  can be remarkably well
  explained by assuming that the decay proceeds first through the
  $\pi^+ \bar{D}^0$ (or $\bar{D}^{*0}$) decay, whose branching ratios are known, and then
  through the subsequent decay of the virtual $\bar{D}^0$ or $\bar{D}^{*0}$  mesons to
  $\Lambda_c^- p$, whose strength are calculated using previously fit
  hadronic parameters.  We then note that the $\Theta_c$ can be
  similarly produced when the virtual $\bar{D}^0$ or $\bar{D}^{*0}$ decay into an
  anti-nucleon and a $\Theta_c$.  Combining the present theoretical
  estimates for the ratio $g_{D N \Lambda_c} / g_{D N \Theta_c} \sim
  13 $ and $g_{D^* N \Theta_c} \sim \frac{1}{3} g_{D N \Theta_c}$,
  we find that the anti-charmed pentaquark $\Theta_c$, which was
  predicted to be bound by several model calculations, can be produced
  via $B^+ \rightarrow \Theta_c \bar{n} \pi^+$, and be observed from
  the $B$-factory experiments through the weak decay of $\Theta_c
  \rightarrow p K^+ \pi^- \pi^- $.

\end{abstract}


\maketitle

%%%%%%%%%%%%%%%%%%%% Text %%%%%%%%%%%%%%%%%%%%%

%\section{Introduction}

The excitement about $\Theta^+$ after its first discovery by the LEPS
Collaboration at SPring-8 \cite{Leps03} and its subsequent
confirmation by several groups, has recently turned into
disappointment and confusion, as increasing number of experiments are
reporting negative results with higher
statistics\cite{hicks,Burkert:2005ft}.  The experimental situation is
similarly discouraging for anti-charmed pentaquark, as the initial
observation by the H1 collaboration at HERA\cite{Aktas:2004qf} has not
been confirmed by subsequent
experiments\cite{Litvintsev:2004yw,Karshon:2004kt,Link:2005ti}.  While
certain processes and energy ranges are ruled out, one can not afford
to give up further refined experimental search, because if a
pentaquark is found, it will not only provide a major and unique
testing ground for QCD dynamics at low energy, but also a basis for
investigating many body properties of QCD at higher density.

Given the experimental situation, one should go back and ask what
kind of insights theoretical considerations can give us in the
search for the $\Theta_c$ or the $\Theta^+$.  In this respect, one
should first note that the theoretical grounds for the heavy and
light pentaquarks are quite different.   There are on going
discussions\cite{Cohen:2003yi, Walliser:2005pi} over the validity
of the original prediction for the mass of the $\Theta^+$ based on
the SU(3) Skyrme model\cite{DPP97}. On the other hand, many
theories consistently predicted that the heavy pentaquark state is
stable and lies below the $DN$ threshold.  The pentaquark with one
heavy anti-quark was first studied in Ref.~\cite{GSR87,Lip87} in a
quark model with color spin interaction, with  flavor spin
interaction in \cite{Stan98}, and Skyrme models \cite{RS93,OPM94},
and recently in
~\cite{Stewart:2004pd,hung04,lutz,Pirjol:2004dw,Cohen:2005bx},
which were motivated by the
diquark-diquark\cite{JW03,Narodetskii:2003ft,Semay:2004jb} and
diquark-triquark\cite{KL03a} picture.   In the constituent quark
model, the existence of a pentaquark state with diquark
configuration, crucially depends on how strong the diquark
correlation is compared to the quark anti-quark correlation when
it recombines into a meson and a nucleon state.  Since both
correlations are effectively inversely proportional to the
constituent quark masses involved, $C \sum_{i>j} \vec{s}_i \cdot
\vec{s_j} \frac{1}{m_i
m_j}$\cite{DeRujula:1975ge,Hiyama:2005cf,Stancu:2004du}, the
attraction is expected to be more effective for pentaquark state
with a heavy anti-quark.    As a simplified example, consider the
pentaquark picture given in ref.\cite{JW03}.  The $ud$ diquark
will form a color anti-triplet, isospin 0 and spin 0 state. Using
the $C$ determined from $M_\Delta-M_N=\frac{3C}{2m_u^2}=290 $ MeV,
one finds an attraction of 290 MeV from the two diquarks.  This
would be identical whether one has a heavy or light anti-quark. On
the other hand, assuming the pentaquark recombines into a nucleon
and a meson, the attraction expected from the diquark correlation
in the spin 1/2 baryon and the quark anti-quark in the spin zero
meson would be $ -\frac{3C}{4m_u^2} -\frac{3C'}{4m_u m_s} = -430$
MeV. Where $C'$ is determined from
$M_{K^*}-M_K=\frac{C'}{m_um_s}=397 $ MeV. If $m_s$ is replaced by
$m_c$, this will become $-240$ MeV, from
$M_{D^*}-M_D=\frac{C_M}{m_um_c}=137 $ MeV.  Therefore, comparing
this to $- 290$ MeV in the pentaquark configuration, one expects a
bound pentaquark state only when the anti-quark is heavy. Such
simple expectations are explicitly bourne out in the constituent
quark model calculations, which predict a more stable pentaquark
configuration when the antiquark becomes
heavy\cite{Stancu:2005jv,Maltman:2004qe,Hiyama:2005cf,Stancu:2004du}.
Similar results are also consistently obtained in the skryme
model\cite{OPM94} and the QCD sum rule calculations\cite{SKL05}.
 In this respect, the negative
experimental result for the heavy pentaquark from the $DN$ or $D^*
N$ final state could be a natural consequence of its stability,
and one should search for it from its weak decay.

There were attempts to search for a stable heavy pentaquark with
strangeness\cite{attempts}, using high energy pion beam on a
nuclear target.  But with no realistic estimate on the production
cross section of the pentaquark, it is difficult to draw any
strong conclusion. Moreover, the QCD sum rule\cite{SKL05} or the
 skyrmion approach predict the heavy pentaquark state to be stable
 only when it has no strangenes\cite{OPM94}.

Therefore, we propose to search for a stable heavy pentaquark without
strangeness, and show that the accumulated data at the $B$-factory
experiments may be sufficient for this purpose.  To be precise, we
will show, that the charmed pentaquark can be produced from the
$B^+\rightarrow \Theta_c \bar{n} \pi^+$ decay, and that with the most
conservative estimate, it can be identified through one of its weak
decays, $\Theta_c \rightarrow p K^+ \pi^- \pi^-$.

%\subsection{Diquark correlation in heavy pentaquark}

%\subsection{Branching ratio}

To build up on a reliable method for estimating the decay, we
first try to understand the baryonic decay mode $B^+ \rightarrow
\Lambda^-_c p \pi^+ $.
%\begin{eqnarray}
%B^+ \rightarrow \Lambda^-_c p \pi^+. \label{b-to-3}
%\end{eqnarray}

\begin{figure}
\centering \epsfig{file=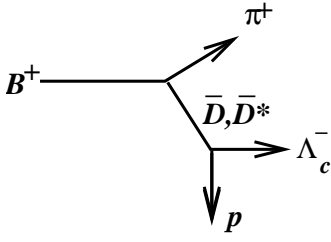, width=.5\hsize} \caption{
Hadronic description for the hadronic decay $ B^+ \rightarrow
\Lambda_c \bar{p} \pi^+$ } \label{3body}
\end{figure}

The branching ratio for this decay is well measured, $(2.1 \pm
0.7)\times 10^{-4}$\cite{Eidelman:2004wy}, which is the largest
among the two and three body baryonic decays. Due to color
suppression factors, $\pi^+$ is most likely produced from the
$W^+$ decay, which changes the $\bar{b}$ quark to a $\bar{c}$
quark. Once this takes place, the remaining $\bar{c} u $, which
has the quantum number of a $\bar{D}^0$ or $\bar{D}^{*0}$, will
subsequently decay into $\Lambda^-_c p$. In a hadronic language,
this process is depicted in Fig.~\ref{3body}. Any other hadronic
diagram is quite unlikely as the $\bar{b}$ quark has to first
decay weakly for it to dissociate into any other baryonic modes.

To calculate the decay rate for the process in Fig.~\ref{3body},
we first estimate the weak decay coupling $G_{B\pi D}$ from the
branching ratio of $B^+\rightarrow \pi^+ D$.  We use the formula
for two body decay with a form factor.
%, to account for the finite size of the hadrons involved.
\begin{eqnarray}
\Gamma_{B^+ \pi^+ D} = \frac{G_{B \pi D}^2}{8 \pi} \frac{ |{\bf
q}|}{m_B^2}|F({\bf q}^2)|^2,  \label{2-body}
\end{eqnarray}
where, $|{\bf
q}|=\frac{1}{2m_B}[(m_B^2-(m_\pi+m_D)^2)(m_B^2-(m_\pi-m_D)^2)]^{\frac{1}{2}}$.
We will use similar hadronic couplings and form factors used
previously, to fit the open charm production from $p-p$ reaction
at low energy\cite{LLK03-2}. The form factors used are of monopole
type given as,
\begin{eqnarray}
F({\bf q}^2)= { \Lambda^2 \over \Lambda^2 + {\bf q}^2},
\end{eqnarray}
where $\bf{q}^2$ is the $D$ meson three-momentum in the rest frame
of the $B$ meson.  We will take $\Lambda=300 $ MeV to best fit the
open histogram to be discussed in Fig. 2. From the measured life
time of the $B^+$ meson of $(1.671 \pm 0.018)\times
10^{-12}s$\cite{Eidelman:2004wy} and the branching ratio for the
decay $B^+\rightarrow \pi^+ \bar{D}^0 (\bar{D}^{*0})$ of $(4.98
\pm 0.29) \times 10^{-3}$ ($(4.6 \pm 0.4) \times 10^{-3}$) , we
find,
\begin{eqnarray}
G_{B \pi D}& = &46 ~{\rm keV},  \nonumber \\
G_{B \pi D^*} & =& 43 ~{\rm keV}. \label{gb}
\end{eqnarray}
Note that these are just phenomenological couplings defined
through Eq.(\ref{2-body}).

\begin{figure}
\centering \epsfig{file=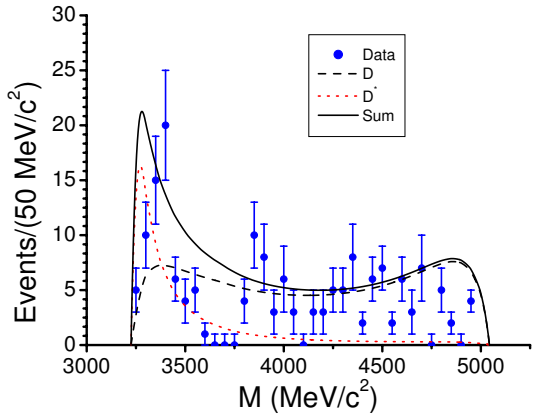, width=1.\hsize} \caption{
Comparison of the experimental data\cite{Abe:2004sr} to the
present calculation on the open histogram of the $B^+\rightarrow
\Lambda_c^- p \pi^+$ decay as a function of $M(\Lambda_c^-p)$,
requiring $M(\Lambda_c^- \pi)>2.6$ GeV$/c^2$ and $M(p \pi^+)>1.6$
GeV$/c^2$. } \label{fig2-c}
\end{figure}

 Once this coupling is given, the three-body
decay rate for the process $ B^+ \rightarrow \Lambda^-_c p \pi^+$,
represented in Fig.~\ref{3body} is given by,
\begin{eqnarray}
\Gamma_{B^+ \rightarrow \Lambda_c \bar{p} \pi^+}=\frac{1}{2^8
\pi^3 m_B^3} \int_{(m_\pi+m_p)^2}^{(m_B-m_\Lambda)^2} d p_{\pi
p}^2 \nonumber
\\ \times \int_{p_{D-min}^2}^{p_{D-max}^2} d p_D^2
|M|^2 |F( {\bf p}_B) F( {\bf p}_D)|^2 , \label{gamma1}
\end{eqnarray}
where
\begin{eqnarray}
{\bf p}_B^2 & =&
\frac{(m_B^2-(p_D+m_\pi)^2)(m_B^2-(p_D-m_\pi)^2)}{4
m_B^2} , \nonumber \\
{\bf p}_D^2 & =&
\frac{(p_D^2-(m_\Lambda+m_p)^2)(p_D^2-(m_\Lambda-m_p)^2)}{4
p_D^2}.
\end{eqnarray}
 The range of integration is given by,
\begin{eqnarray}
p_{D-min,max}^2 & =&  (E_p^*+E_\Lambda^*)^2 \nonumber \\
 & -& \bigg(  \sqrt{ E_p^{*2}-m_p^2}\pm
\sqrt{E_\Lambda^{*2}-m_\Lambda^2} \bigg)^2,
\end{eqnarray}
where
\begin{eqnarray}
E_p^* & = & \frac{1}{2p_{\pi p}}(p_{\pi p}^2-m_\pi^2+m_p^2) \nonumber \\
E_\Lambda^* & = & \frac{1}{2p_{\pi p}}(m_B^2-p_{\pi
p}^2-m_\Lambda^2).
\end{eqnarray}
 For $D$ intermediate
state, the matrix element is given as
\begin{eqnarray}
|M_D|^2  = 2 g_{D P \Lambda_c}^2 G_{B \pi D}^2
{p_D^2-(m_p+m_\Lambda)^2 \over  (p_D^2-m_D^2)^2}
\end{eqnarray}
and for $D^*$ as,
\begin{eqnarray} |M_{D^*}|^2  =
\frac{4 g_{D^* P \Lambda_c}^2 G_{B \pi
D^*}^2}{(2m_B^2+2m_\pi^2-m_{D^*}^2)}  { 1 \over
(p_D^2-m_{D^{*}}^2)^2}  \nonumber \\
 \times \bigg(2(p_{\pi
p}^2+\frac{p_D^2}{2}-\frac{m_p^2}{2}-\frac{m_\Lambda^2}{2}-m_\pi^2)\nonumber
\\ \times (m_B^2+
\frac{m_\Lambda^2}{2}+\frac{m_p^2}{2}-\frac{p_{\pi
p}^2}{2}-\frac{p_D^2}{2}) \nonumber \\
+(m_B^2+m_\pi^2-\frac{p_D^2}{2})((m_p+m_\Lambda)^2-p_D^2) \bigg),
\end{eqnarray}
where for simplicity, we have assumed  here and in obtaining
Eq.(\ref{gb}) that the polarization sums for the $D^*$ are just
proportional to $g_{\mu \nu}$.  Using $g_{D P \Lambda_c}=13.5$ and
$g_{D^* P \Lambda_c}=-4$~\cite{LLK03-2} and Eq.(\ref{gb}), we
find,
\begin{eqnarray}
\Gamma_{B^+ \rightarrow \Lambda^-_c p
\pi^+}/\Gamma_{B^+}=(2.05+0.51) \times 10^{-4},  \label{decay3}
\end{eqnarray}
where the first(second) number come the $D$($D^*$) intermediate
state.  This lies within  the experimental measurement of $(2.1
\pm 0.7) \times 10^{-4}$.
%, as we neglected contributions from intermediate baryon
%states\cite{Hychen}.  However, the present contributions,
%comprises a dominant part of the decay.
In Fig \ref{fig2-c}, we compare our calculation for the open
histogram as a function of $p_D=M(\Lambda_c^-p)$ to the
experimental result in ref.\cite{Abe:2004sr}.  The theoretical
event rates are obtained from our differential formula in
Eq.(\ref{gamma1}) multiplied by the total number of 152 million
$B^+$, used in ref.\cite{Abe:2004sr}.
%, and an arbitrary normalization factor of 0.5.
The solid line is the sum of the contributions from $D$ and $D^*$.
Noting that the theoretical form factors and couplings were
similar to those used in a totally different
reaction\cite{LLK03-2}, the agreement is quite remarkable.
%Given the large experimental uncertainties, we do not try to refit the
%coupling to account for the factor of 0.5 needed to fit the
%normalization at this stage.
\begin{figure}
\centering \epsfig{file=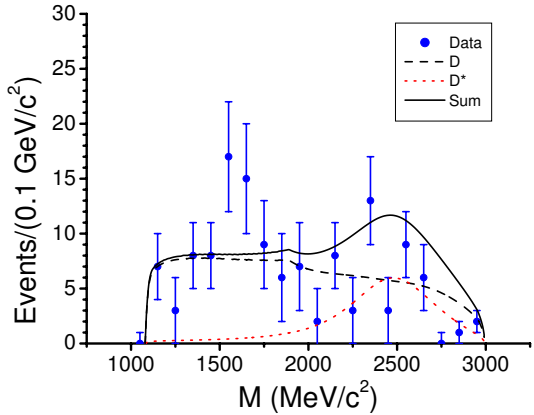, width=1.\hsize} \caption{
Similar to Fig. \ref{fig2-c}, but as a function of $M(p \pi^+)$,
requiring $M(\Lambda_c^- \pi)>2.6$ GeV$/c^2$ and $M(\Lambda_c^- p
)>3.5$ GeV$/c^2$. } \label{fig2-b}
\end{figure}
In Fig. \ref{fig2-b}, we also show the histogram in $p_{\pi
p}=M(\pi^+ p)$. Apart from the peak in 1.6 GeV\cite{Hychen}, which
should come from the $\Delta(1600)$ intermediate state that we did
not include, the other structure and magnitude is again well
reproduced.  We conclude from our fit that the $D$ and $D^*$
intermediate state contributions are important part of the
baryonic decay  $B^+ \rightarrow \Lambda^-_c p \pi^+$, and
explains much of the detailed histogram of its decay.   We note
that such contributions have to be subtracted out from the data
before extracting any information about baryon resonance mediated
decay\cite{Abe:2004sr}.

%\subsection{Branching ratio for $\Theta_c$ production}

\begin{figure}
\centering \epsfig{file=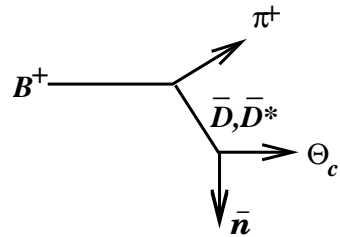, width=.5\hsize} \caption{
Hadronic description for the hadronic decay $ B^+ \rightarrow
\Theta_c \bar{n} \pi^+$ } \label{thetac}
\end{figure}

Let us now consider a process where a $\Theta_c$ can be produced.
The process is shown in Fig.~\ref{thetac}.
We use the same formula as in Eq.(\ref{gamma1}) with masses
replaced to those shown in Fig. \ref{thetac}.  We take the
$m_\Theta = 2800$  MeV, which is slightly below the $DN$
threshold, and assume it to have spin 1/2 and positive parity. To
estimate the coupling $g_{DN \Theta_c }$, we use an analogy to the
$\Theta^+$.

The particle data book puts the width of $\Theta^+$ to be $0.9 \pm
0.3 $ MeV\cite{Eidelman:2004wy}.  Such a small width is necessary
if the existence of the resonance is to be consistent with the
previous $KN$ scattering data.  Moreover, all the experiments
reporting positive signal quote the width to be smaller than their
experimental resolution, which are typically of 10 MeV.  Now
assuming the width of $\Theta^+$ to be 1 MeV, which is dominated
by its $KN$ decay, one finds that the coupling to be
$g_{KN\Theta^+}=1$\cite{Oh:2003gj}.  Noting that $g_{D P
\Lambda_c}$ is estimated to be similar in magnitude to
$g_{KN\Lambda}$\cite{LLK03-2}, we will also take
$g_{DN\Theta_c}=1$.  Such a small coupling is also expected if the
pentaquark wave function is composed of strongly correlated
diquarks with small spatial overlap with the $D N$
states\cite{JW03}.  Moreover, we will take
$g_{DN\Theta_c}/g_{D^*N\Theta_c}=g_{DP\Lambda_c}/
g_{D^*P\Lambda_c}\sim1/3$.   With this coupling, we find the
branching ratio for $ B^+ \rightarrow \Theta_c \bar{n} \pi^+$ to
be $14.4\times 10^{-7}$, which roughly comes from
$(g_{DN\Theta_c}/g_{DP\Lambda_c})^2\times \Gamma_{B^+ \rightarrow
\Lambda_c^-p\pi}/\Gamma_{B^+}$.

%\subsection{Experimental Prospects}

Once $\Theta_c$ is produced, and if it is unbound, it can decay
into either $D^- p$ and $\bar{D}^0 n$, and be directly observed.
However, if it is bound, it will only be observed via weak decays.
The dominant decay would be through $\Theta_c \rightarrow p K^+
\pi^- \pi^-$.  To estimate this branching ratio, we assume that it
is   similar to that of $D^- \rightarrow K^+ \pi^- \pi^-$, which
has a branching ratio of $(9.2\pm 0.6)$\%\cite{Eidelman:2004wy}.

To account for experimental acceptance and efficiency, we take
70\% as a rough estimate of track-finding efficiency including
particle identification, for each charged particle. Then, the
total efficiency to correctly find all 4 charged tracks for a
$\Theta_c$ decay would be $(0.7)^4$. Combining the two $B$-factory
experiments, Belle and BABAR, we are close to accumulating $10^9$
$B^+ / B^-$ pairs.  Therefore, the total number of expected events
for $B^+ \rightarrow \pi^+ \bar{n} \Theta_c$ and subsequently
$\Theta_c \rightarrow p K^+ \pi^- \pi^- $, would be,
\begin{eqnarray}
(10^9)(14.4 \times 10^{-7})(0.092)(0.7)^4 =32 ~~{\rm events}.
\label{final-number}
\end{eqnarray}

In a sense, this can be regarded as a lower limit, since the
contribution from other possible production processes have been
neglected and a very conservative estimate for $g_{DN \Theta_c}$
has been taken. The main uncertainty of this number comes from our
uncertainty in the overall fit of Fig 2 and Fig. 3.  If
$m_{\Theta_c}=3100 $MeV and unbound, the branching ratio will only
change slightly to $15.5\times 10^{-7}$, and its existence could
be observed through strong decay into $DN$ or $D^*N$ final states
with event rates larger than that given in
Eq.(\ref{final-number}), depending on what additional final states
will be used to identify the on shell D meson.

Considering the fact that each $B$-Factory experiment will
accumulate at least $1\times 10^8$ $B^+$'s every year, adding the
already accumulated data of about $0.5 \times 10^9$, the prospect
of observing a charmed pentaquark state in this channel from
analyzing the existing and upcoming data at each factory is quite
promising.

%\subsection{Summary}

We have shown that the baryonic decay modes of $B^+$ can be
sensibly estimated with previously determined hadronic parameters.
Starting from such methods and previous estimates on the coupling
of the pentaquark to the $DN$ states, we find that the pentaquark
can be produced in the baryonic decay of the $B^+$. Estimates show
that in both cases, where the pentaquark is unbound or bound, the
pentaquark can be observed realistically through the $DN$ or weak
decay final states respectively, from the accumulated data at
B-Factories. Previous searches on the charmed pentaquark were not
combined with a realistic estimate of the cross section, and
therefore it was not clear what to conclude even from a null
result.   Since we present a definite lower bound on the counting
rate with which the pentaquark should be observed, the
experimental search would be able to provide a final conclusion on
the existence of the charmed pentaquark.

We are grateful to A. Hosaka and H.Y. Cheng for useful discussions
and information.  The work of S.H.L was supported by Korea
research foundation under grant number C00116. The work of Y.K.
was supported by Korea research foundation under grant number
KRF-2005-070-C00030.

\end{document}